\documentclass[aps,jap,twocolumn]{revtex4-1}
\usepackage{dcolumn}
\usepackage{bm}

\usepackage[tight,TABTOPCAP]{subfigure}
\usepackage{amsmath,amsfonts} 
\usepackage{amssymb}

\usepackage{graphicx}

\newcommand{\ud}{\mathop{\mathrm{{}d}}\mathopen{}}
\newcommand{\be}{\begin{equation}}
\newcommand{\ee}{\end{equation}}
\newcommand{\demi}{\frac{1}{2}}

\newcommand{\re}{\mbox{Re}}
\newcommand{\norm}[1]{\left|#1\right|}

\begin{document}

\title{Non-reciprocal Light-harvesting Nanoantennae \\ Made by Nature}

\author{Julian Juhi-Lian Ting}
\homepage{http://amazon.com/author/julianting}
\email{juhilian@gmail.com}
\affiliation{De-Font Research Institute, Taichung 40344, Taiwan, R.O.C.}
\date{\today}

\begin{abstract}

Most of our current understanding of mechanisms of photosynthesis comes from spectroscopy.
However, classical definition of radio-antenna can be extended to optical regime to discuss the function of light-harvesting antennae.
Further to our previously proposed model of a loop antenna we provide several more physical explanations on considering the non-reciprocal properties of the light harvesters of bacteria.
We explained
the function of the non-heme iron at the reaction center,  
and presented reasons for each module of the light harvester being composed of one carotenoid, two short $\alpha$-helical polypeptides and three bacteriochlorophylls;
we explained also the toroidal shape of the light harvester, 
the upper bound of the characteristic length of the light harvester, 
the functional role played by the long-lasting spectrometric signal observed, and the photon anti-bunching observed. 
Based on these analyses, two mechanisms might be used by radiation-durable bacteria, {\it Deinococcus radiodurans}; 
and the non-reciprocity of an archaeon, {\it Haloquadratum walsbyi}, are analyzed.
The physical lessons involved are useful for designing artificial light harvesters, optical sensors, wireless power chargers, 
passive super-Planckian heat radiators, photocatalytic hydrogen generators, and radiation protective cloaks.
In particular it can predict what kind of particles should be used to separate sunlight into a photovoltaically and thermally useful range to enhance the efficiency of solar cells.

\end{abstract}


\maketitle

\section{Introduction}

There are three plus one types of photo-autotrophs on earth, i.e.
plants, algae and bacteria plus archaea.
Archaea use a photosynthetic mechanism quite different from the others.
The photosynthesis units of plants and algae are much more complicate than bacteria,
which is a subject traditionally studied by chemists and biochemists~\citep{Voet2011,Berg2015}.
A major task in this field was trying to understand the molecular structures.
Because of the inadequate instrumental resolution, 
people struggled for decades to know the structure of the light-harvesting antenna.
Without precise structures, investigators tried to guess the content within the black box.


Since about 1995, satisfactory pictures of the bacterial light-harvesting (LH) systems have been obtained~\citep{Kuhlbrandt1995a}.
Both the inner antenna (LH1) and the outer antenna (LH2) have a toroidal shape and are composed of the same modules. 
Each module contains one carotenoid two short $\alpha$-helical polypeptides, and three bacteriochlorophylls. 
The exact numbers of modules involved for both complexes are variable, as shown in Table \ref{more}. 
The outer antenna, LH2, is smaller and consists of nine units for 
{\it Rhodopseudomonas acidophila} (1NKZ)~\citep{Papiz2003}
; 
the inner antenna, LH1,  is larger, as it contains the reaction center (RC), and has 17 units, 
for {\it Blastochloris viridis} (6ET5) 
~\citep{Qian2018}.
The exact dimension of each molecule can be read by software such as Jmol or PyMOL from a .cif or .pdb files in the PDB (protein data bank http://www.rcsb.org/pdb/).
Symmetries are notably unaltered even under strained conditions for LH3, i.e., a variant of LH2~\citep{McLuskey2001}, which indicate their importance.


Two further subtleties in the structure of LH1 are important. 
First, the RC contained in the LH1 has a non-heme iron.
Second, for some species the ring of the LH1-RC complex has an opening,
perhaps with a polypeptide termed PufX or W present over that region ~\citep{Holden-Dye2008,Olsen2017}.

Molecular data show that photosystems I/II of algae and plants likely evolved from the photosystems of green-sulfur bacteria; 
there are hence many analogous functions and similar structures, except with more sophisticated material such as 
not only non-heme iron but even manganese ($Mn$) are presented at their RC~\citep{Mueh2013,Semenov2011}.


\begin{table}[tbp]
\begin{tabular}{p{0.6\columnwidth}ccc}
\hline 
protein & PDB ID & symmetry & cartoon \\
\hline \\

LH1-RC from {\it Blastochloris viridis} ~\citep{Qian2018} & 6ET5 & C17 &
\begin{minipage}[c]{0.15\columnwidth}\includegraphics[width=\columnwidth, angle=0]{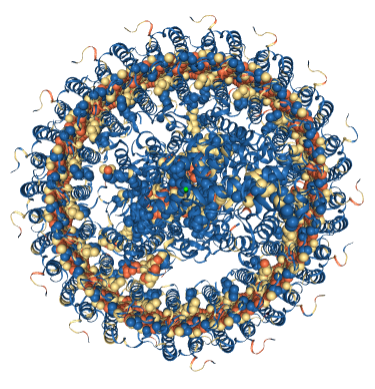}\end{minipage}  \\

LH1-RC  from {\it Rhodopseudomonas palustris} ~\citep{Roszak2003}              & 1PYH &  C2 &
\begin{minipage}[c]{0.15\columnwidth}\includegraphics[width=\columnwidth, angle=0]{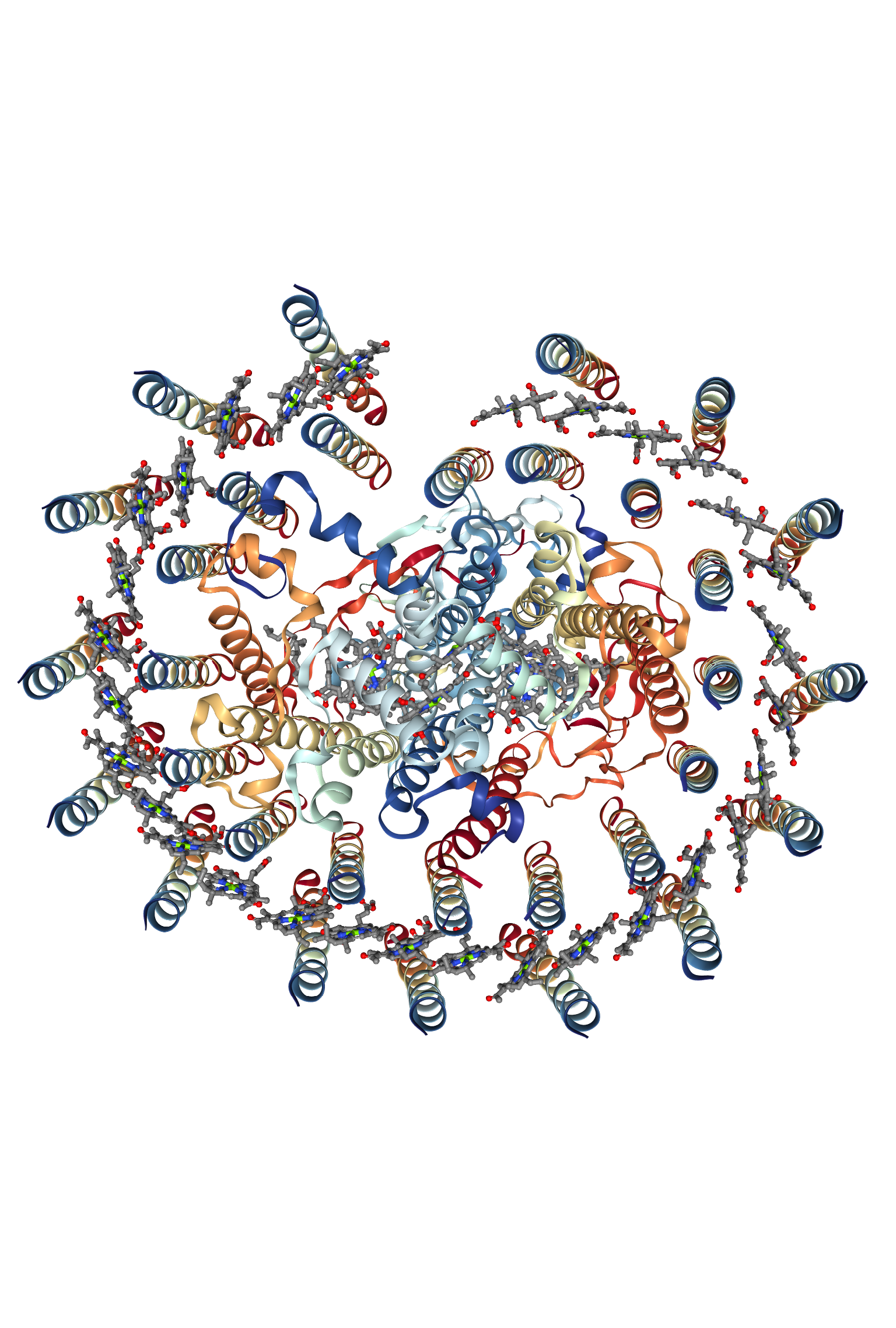}\end{minipage}  \\

LH2 B800-850 from {\it Rhodopseudomonas acidophila} ~\citep{Papiz2003}              & 1NKZ & C9 &
\begin{minipage}[c]{0.15\columnwidth}\includegraphics[width=\columnwidth, angle=0]{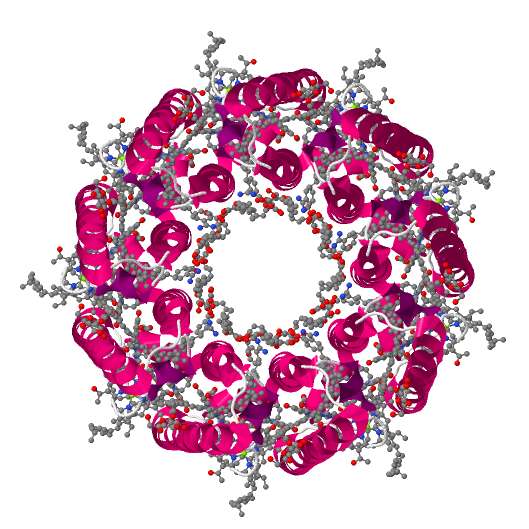}\end{minipage}  \\

LH2 B800-850 from {\it Rhodospirillum molischianum} ~\citep{Koepke1996}              & 1LGH & C8 &
\begin{minipage}[c]{0.15\columnwidth}\includegraphics[width=\columnwidth, angle=0]{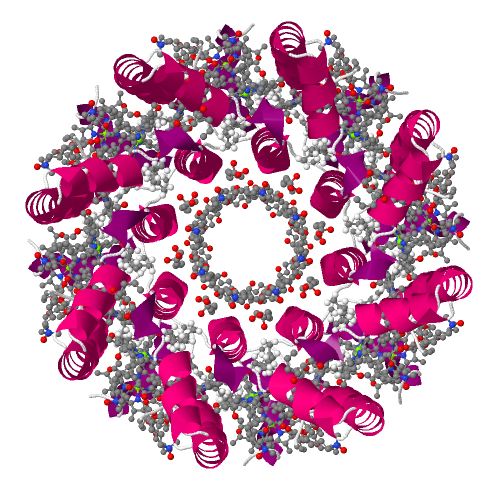}\end{minipage}  \\

\\
\hline 
\end{tabular}
\caption{Various light harvesters with structural symmetry.
}
\label{more}
\end{table}

On the other hand, we do not know much about the structure of archaea photosynthetic unit yet,
which were classified as separate to bacteria only in the 1970s.
Amongst them {\it Haloquadratum walsbyi} (Hqr. in genus abbreviation) inhabiting in hypersaline environments was discovered by Anthony Walsby in 1980~\citep{Walsby1980,Walsby2005}.
It grows by using simple carbon-containing compounds leaked from other microbes in the salty lakes as energy sources, but can also convert light into usable energy.
Its cell, which is the basic photosynthetic unit in compare to the LH of bacteria, 
is square-like shape with sharp corners and edges measure about $2 - 5~ \mu m$ while its thickness is only around $0.1 - 0.2~ \mu m$,
unlike the spherical or cylindrical shape of many other organisms,
with a cross-shape dark line on its surface.
The cell wall consists of bacteriorhodopsin and surface layer (S-layer) glycoproteins~\citep{Bolhuis2006}.

Several kinds of bacteriorhodopsin have been analyzed with x-ray crystallography.
5ITE, 5ITC, 5KKH, and 4QI1 are in trimeric form, whereas 4QID, 4WAV are in dimeric form.
They occupy up to 50\% area of the cell surface of the archaea. 
These bacteriorhodopsins form a hexagonal lattice composed of three identical protein chains, 
each rotated by 120 degrees relative to the others.
They are used as light-gated proton pump to convert captured solar energy into ATP, which
is different from all other phototrophic systems in bacteria, algae, and plants that use chlorophylls or bacteriochlorophylls. 
Three alternative bacteriorhodopsins carry out photosynthetic growth; two are proton pumps and the other is a chloride pump.
The absorption spectrum has a peak at around $568~nm$, which is at the green light range
and with a photon energy of about $2.18~eV$ in vacuum~\citep{Ohtani1995,Schenkl2007}.
Accordingly to Walsby, the shape of the cell is determined by the pattern in which the cell envelope particles assembled.

Even after a knowledge of the structure, questions remain.
Why has a light harvester of bacteria the form of a tambourine-like shape instead of a spherical shape or a serpentine shape, for instance?
Is the number of modules involved important? What does these numbers implied?
Why does each module contains one carotenoid two short $\alpha$-helical polypeptides, and three bacteriochlorophylls?
Are these numbers important?
Why is the surface area to volume ratio of Hqr. so large?
Is the characteristic length of the cell important?
Could we grow a larger cell?
How and why does the cell had a square shape? 

The standard explanations for photosynthesis are given in terms of chemistry~\citep{Voet2011,Berg2015}, 
partly because the method used to study the paths and time scales of transfer of excitation energy is, in general, spectrometry.
There are few physical explanations, as defined by  Gruner {et al.}~\citep{Gruner1995}; 
people tried to mimic the nature from chemical points of view.
For instances, Harriman proposed an artificial light harvester design at molecular level based upon chemical requirements~\citep{Harriman2015}.
A theoretical energy model is constructed and calibrated against spectroscopic data~\citep{Chenu2017}, 
which is more or less a kind of data fitting. 
We tried to incorporate structural information in a model by calculating a simplified LH1 model based upon the chemical rate equations for the shape acquired,
but advanced no further than others~\citep{Ting1999b}.

But physical explanations are possible as we extended classical definition of radio-antenna to 
optical regime~\citep{Ting2018,Ting2019}.
In two previous papers, we explained the reasons for
\begin{itemize}
\item the function of the notch at the light harvester, 
\item the function of the PufX/W presented at the notch,
\item the function of the special pair,
\item the dimerization mechanism under intensive light,
\item the shape of the light harvester must not be spherical and the cross section of the light harvester must not be circular,
\item the use of dielectrics instead of conductors to make the light harvester,
\item a mechanism to prevent damage from excess sunlight,
\item a mechanism to achieve dual-band radiation spectrum, and
\item reasons for the modular design of the light harvester.
\end{itemize}
In the present paper, we consider the non-reciprocity of the bacteria light-harvesting antennae, 
which can further explain the function of the non-heme iron at the reaction center, 
the toroidal shape of the light harvester,
some spectrometric observations and much more.
After analyzing its reciprocity we extend our analysis to archaea in section \ref{archaean}, 
and list out some possible applications
along with two interpretations for a radio durable bacteria in section \ref{application}.
We begin with a review of the loop antenna model:

\begin{figure}[tb]
\begin{center}

\mbox{
\subfigure[]{\includegraphics[width=0.5\columnwidth, angle=0]{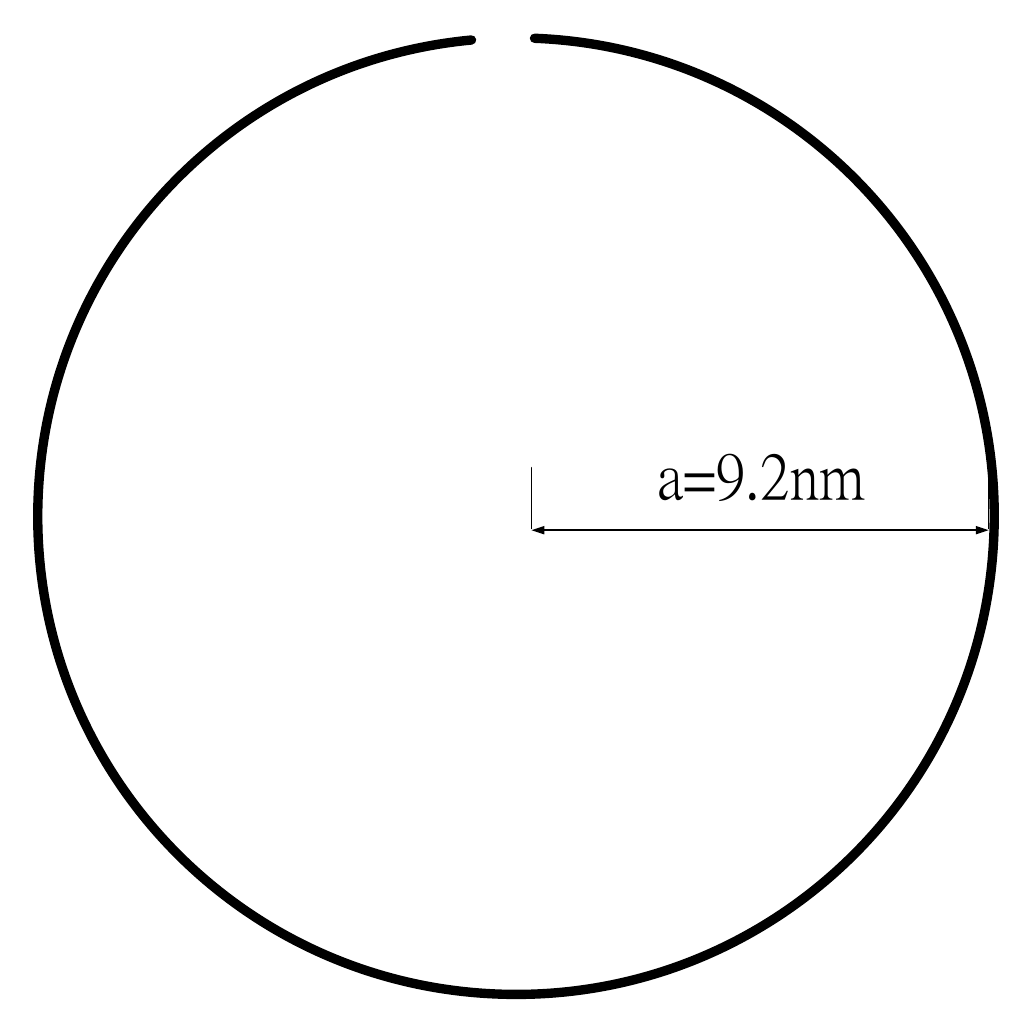}}
\subfigure[]{\includegraphics[width=0.5\columnwidth, angle=0]{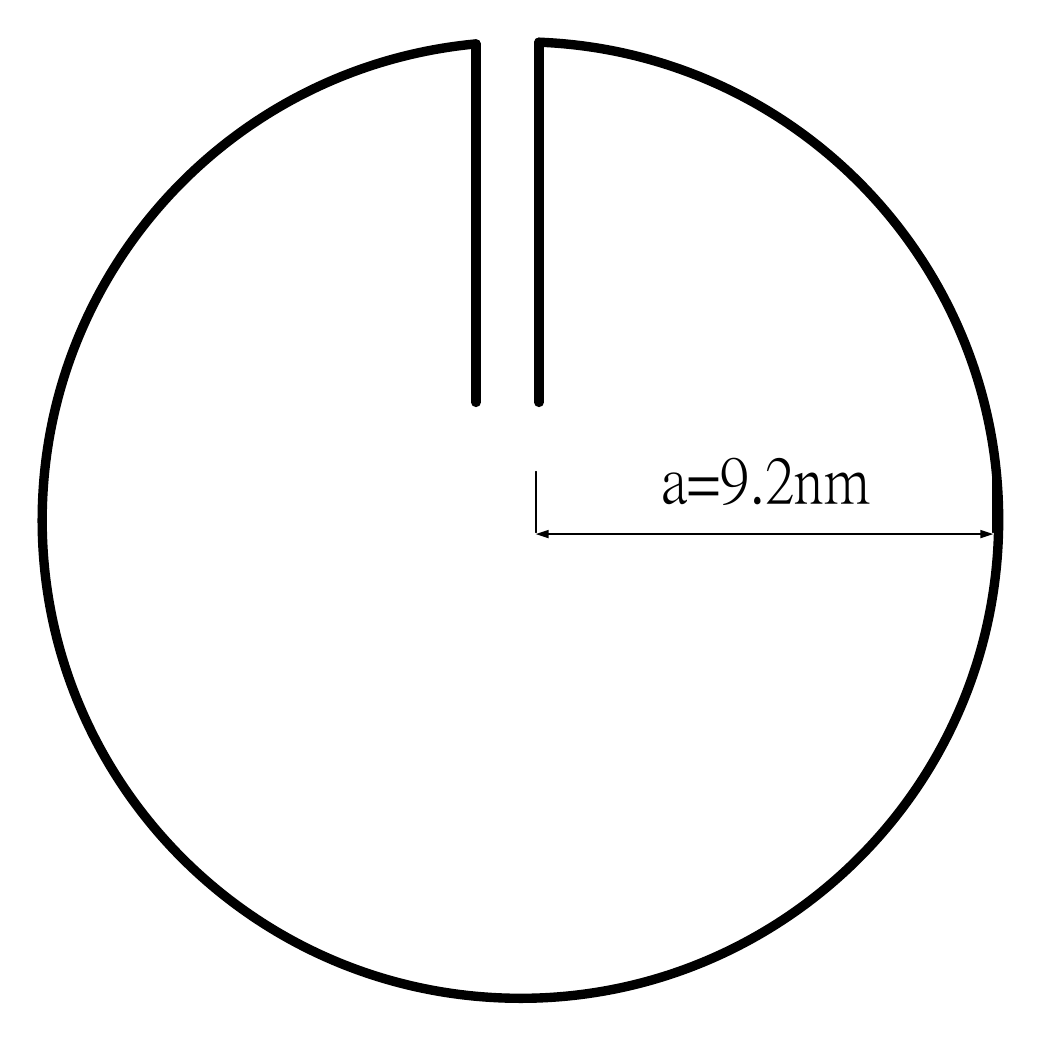}}
}
\caption{Antennae of two simplified shapes. 
(a) is called a (split-ring) loop antenna whereas (b) is called a loop antenna with line feed. 
The radius marked is a typical average of inner diameter and outer diameter measured.
Although resembling the special pair in the RC, the feed line modifies
the resonant frequency only slightly.
To adjust the resonance frequency, the opening can be filled with other
dielectric.
}
\label{models}
\end{center}
\end{figure}

\section{Loop Antennae}
\label{loop}

How should we model the light harvester?  
We want the simplest model which captures all, if possible, essential future instead of including every atoms.
But what are the essential details of the light harvester that we cannot ignore?
The most important parameter of an antenna is its geometry; 
the material properties are immaterial.
The ring shape is apparently one of the essential futures.
Symmetry as indicated is part of that.
There have indeed such shaped antennae considered in classical radio frequency antenna theories.
Figure \ref{models} (a) is typically solved analytically, 
whereas figure \ref{models} (b), although resembling LH1 better, has a resonant frequency only slightly modified from that of figure \ref{models} (a).
We hence consider only figure \ref{models} (a).
The notch is essential for $LH1$, as the received energy must be taken from some point.
The {\it Rhodopseudomonas palustris} molecule (1PYH), shown in the table, clearly also has a notch;
while the opening for {\it Blastochloris viridis} (6ET5) is less obvious.
The opening can be filled with other material to adjust the
resonant frequency in engineering~\citep{Monticone2017}, 
for which purpose a polypeptide called PufX/W is present in LH1~\citep{Swainsbury2017}.
The physical origins of PufX/W and of the notch differ; one does not imply the other. 


Loop antennae are classified into two categories: 
If the antenna has a radius larger than the wavelength of operation, 
it is called a resonant-loop antenna;  otherwise it is called a small-loop antenna~\citep{Balanis2016}.
Because the length scale of free radiation, $850~nm$, 
is much larger than the characteristic size of the antenna, $9.2 ~ nm$, the light-harvesting antennae belong to the category of  small-loop antennae, 
which have small radiation resistances than their loss resistances and serve mainly as receivers.
Similar arguments lead to the small cross-section for the generation of excitons which is a fundamental process for solar energy conversion.
Such dimension is also much smaller than the thermal wavelength, $\hbar c/k_B T \approx 7.5~ \mu m$ at room temperature, hence fulfill the requirement of super-Planckian radiation~\citep{Thompson2018}.
A small-loop antenna is equivalent to an infinitesimal magnetic dipole whose axis is perpendicular to the plane of the loop~\citep{Pendry1999}.

To arrive at the electromagnetic properties of such an antenna there are a simple way and a complete way. 
We begin with the complete way.

\begin{figure}[tb]
\centering
\mbox{
\subfigure[]{\includegraphics[width=0.5\columnwidth, angle=0]{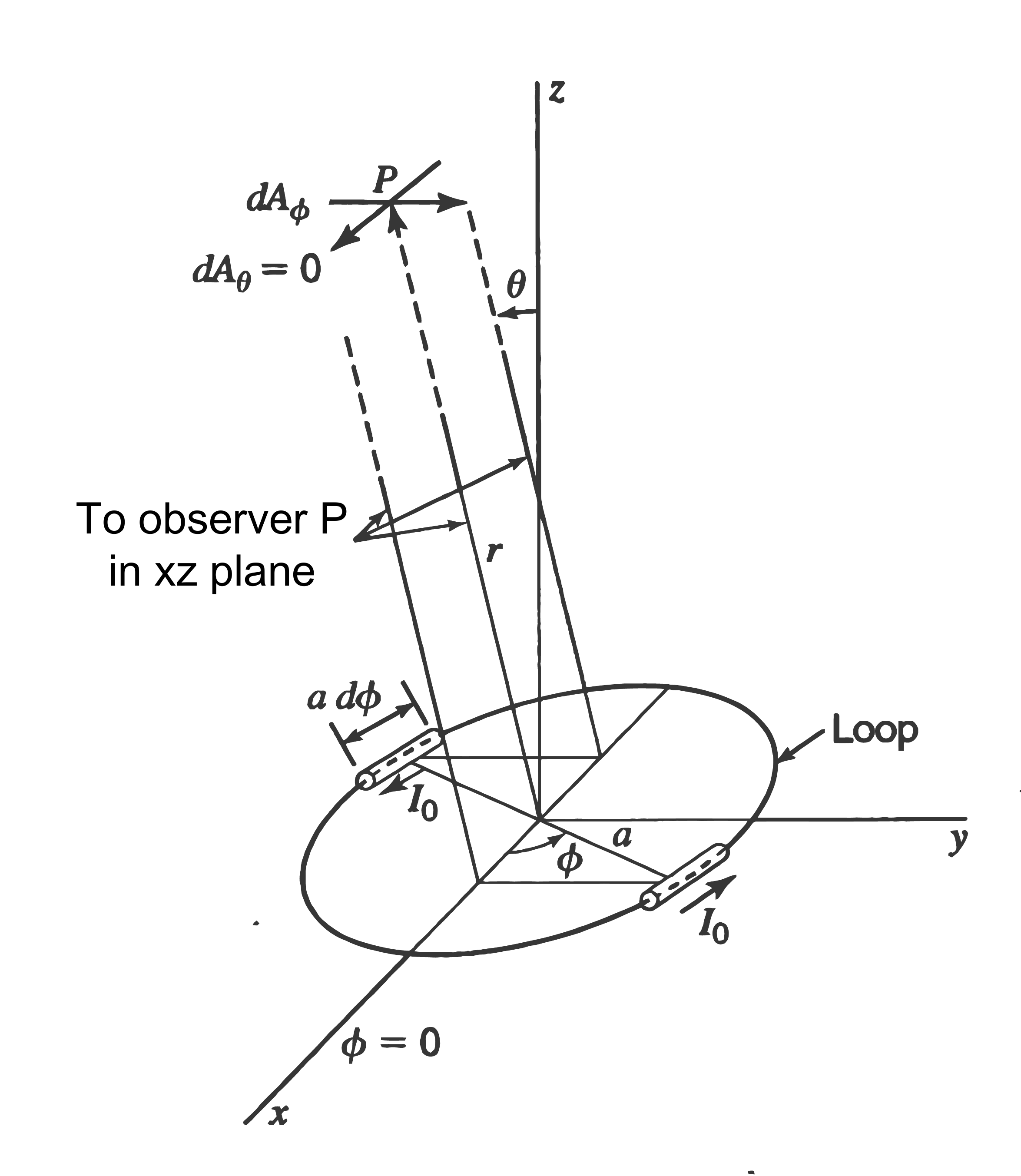}}
\subfigure[]{\includegraphics[width=0.5\columnwidth, angle=0]{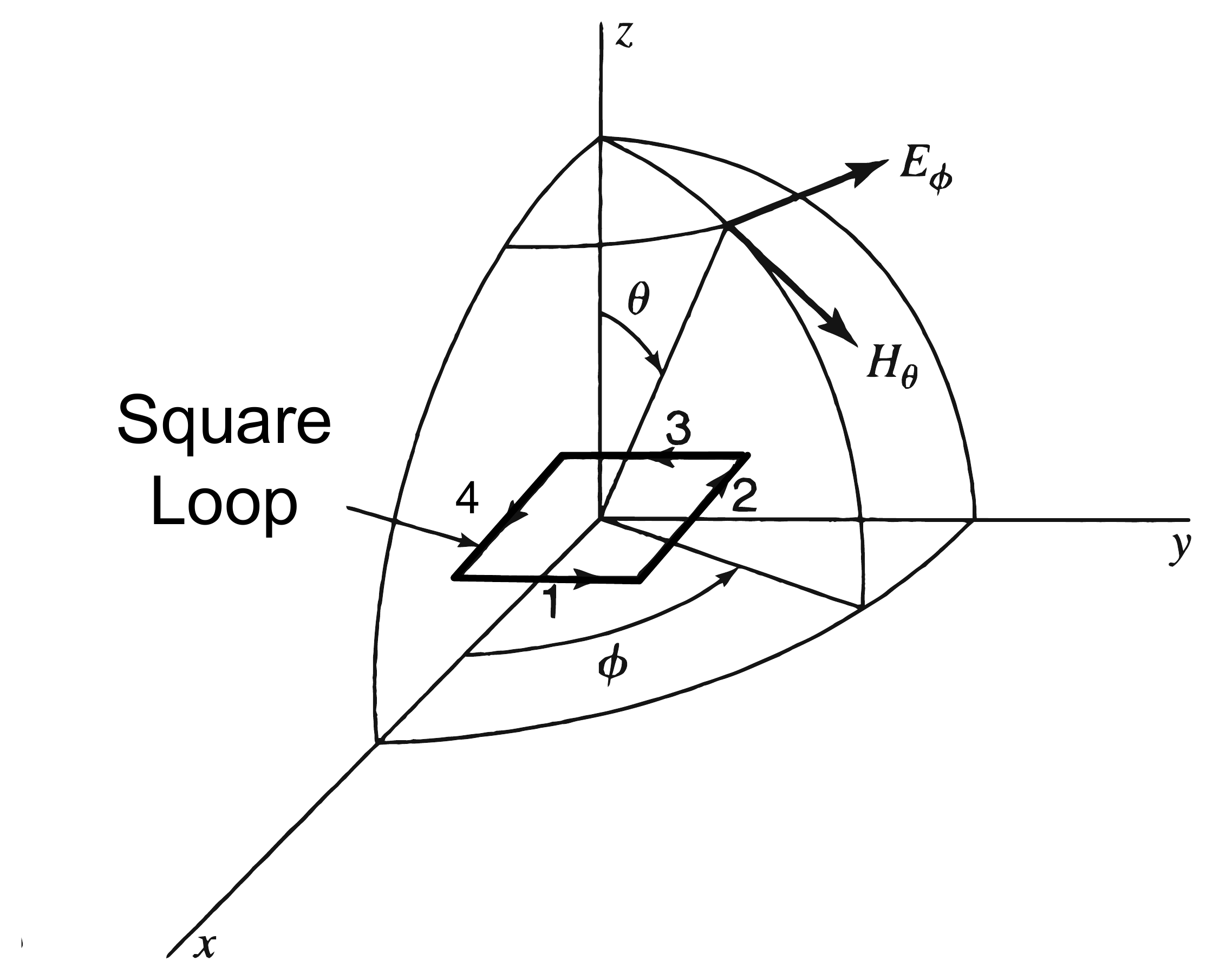}}
}
\caption{The coordinate system uses.}
\label{geometry}
\end{figure}

Let the radius of the loop located at the origin be $a$, and the plane of the loop be $x-y$; 
let the angle from the $x-$axis be $\phi$. If current $I$ around the loop is uniform and in phase,
the only component of the vector potential is $A_\phi$, as shown in Figure \ref{geometry} (a).
The infinitesimal value of $A_\phi$ at a point distant $r$ away from the loop caused by two diametrically opposed infinitesimal dipoles is
\be
\ud A_\phi = \frac{\mu \ud M}{4 \pi r} \,,
\ee
in which $\ud M = 2 j [I] a \cos \phi [\sin (2 \pi a \cos \phi \sin \theta / \lambda) ] \ud \phi$,
$\theta$ is the angle relative to the vertical axis through the center of the loop, 
and
$[I]= I_0 \exp{\{j \omega [t-(r/c)]\}}$ is the retarded current on the loop with $I_0$ being its maximum value.
After integration we obtain
\be
A_\phi = \frac{j \mu [I] a}{2r} J_1 (\frac{2 \pi a \sin \theta}{\lambda} ) \,,
\ee
in which $J_1$ is a Bessel function of first order.

We consider the far-field effects as the source of sunlight is remote. 
The far electric field of the loop has only a $\phi$-component $E_\phi = -j \omega A_\phi$ that is in the plane of the loop. Therefore,
\be
E_\phi = \frac{120 \pi^2 a [I]}{\lambda r} J_1 (\frac{2 \pi a \sin \theta}{ \lambda}) \,.
\label{complete1}
\ee
The corresponding magnetic field in free space reads
\be
H_\theta = \frac{\pi a [I]}{\lambda r} J_1 (\frac{2 \pi a \sin \theta}{ \lambda}) \,.
\label{complete}
\ee

Four short linear dipoles suffice to approximate the loop whilst still preserving the symmetry of the problem, which is the second method.

Let the length of the dipoles be $d$, and
the area of the antenna, which is commonly called the aperture of the antenna, be $A$,
as shown in Figure \ref{geometry} (b). 
Hence
\be
d^2 = \pi a^2 \equiv A
\ee
At the far field
\be
E_\phi = \frac{120 \pi^2 [I] \sin \theta }{r} \frac{A}{\lambda^2} \,.
\label{simple1}
\ee
$A / \lambda^2$ is called the dimensionless aperture.
The magnetic field is obtained on dividing by the intrinsic impedance of the medium, i.e. 
\be
H_\theta = \frac{E_\phi}{120 \pi} = \frac{\pi [I] \sin \theta }{r} \frac{A}{\lambda^2}
\label{simple}
\ee
in a vacuum.
Eq. \eqref{simple} is a special case of Eq. \eqref{complete}, 
just as 
Eq. \eqref{simple1} is a special case of Eq. \eqref{complete1}, 
as for small arguments, $J_1 (x) \approx x/2$.

The (radiation or receiving) resistance at the loop terminals can be obtained from
\be
P = \frac{I_0^2}{2} R
\ee
in which $I_0$ is the maximum current on the loop and $R$ is the resistance.
Integrating the Poynting vector
\be
S = \demi \norm{H}^2 \re Z
\ee
over a large sphere, in which $Z$ is the impedance of the medium, the total power, $P$, is obtained.
The resistance for a small-loop antenna is therefore proportional to $1/\lambda^4$.


The analyses above are direct consequences of Maxwell's equations (in vacuum) that imply no restriction on the range of frequency applicable~\citep{Frimmer2015}.
Symmetry breaking of the electric field in space facilitates the antenna to radiate~\citep{Sinha2015}, 
which is the third reason for the opening at the LH1-RC complex and explaining the non-circular shape of the cross section observed.
We did not assume material property, such as polarizabilities, unlike previous authors~\citep{Zimanyi2010}.
Penetration depth is not an issue, as people working in optical nanoantennae considered, 
because the wire of our idealized antnnae is infinitesimally thin~\citep{Novotny2017}.
An experiment is proposed based upon the above analyses~\citep{Ting2018}.

\section{Non-reciprocal Antennae}

A regular antenna both receives and emits, because Maxwell's equations are symmetric with respect to time reversal~\citep{Collin2001},
but a light harvester that functions solely as a receiver must receive much better than it emits.
What we seek is more than an antenna working with diodes, 
which is artificial and is known as rectenna since 1960s~\citep{Brown1984},
but an optical counterpart of a duckbill check valve in fluid dynamics, which is a naturally designed passive device that exists in a human heart.

As no modal or polarization properties of sunlight are known,
we require mechanisms to break the (Lorentz) time-reversal symmetry~\citep{Jalas2013a}.
There are several possibilities to make such an optical check valve:
\begin{itemize}
\item Faraday rotator~\citep{Faraday1839,Saleh2007};
\item duplexer~\citep{Lira2012,Tzuang2014,Kim2015};
\item nonlinearity (metamaterial) ~\citep{Lepri2011,Mahmoud2015,Krause2008a,Poulton2012a,Tocci1995b,Gallo2001c}; and
\item anapole (toroidal dipole) ~\citep{Huang2013a,Miroshnichenko2015}.
\end{itemize}
All four mechanisms are well known to the scientific and engineering communities.
\begin{itemize}
\item The first mechanism, gyrotropic materials which use Faraday effect, requires an externally applied magnetic field.
All LH1-RC complexes are equipped with a non-heme iron at the RC, 
which could provide the required field.
The iron is not bound to any protein and can be exchanged with zinc ($Zn^{2+}$) or manganese ($Mn^{2+}$), 
just like a pearl in a mussel's mouth that chemists call "coordinated"~\citep{Lawrance2010}.
The promising perovskite material for solar cells also have a lead ion ($Pb^{2+}$) at its center~\citep{Yang2017}.
The role played by the non-heme iron has long been questioned, 
if not unknown to biochemists~\citep{Hermes2006}.
Biologists know that it serves as a source or sink of electrons during electron transfer or redox (reduction-oxidation) chemistry~\citep{Nordlund2011}, 
without recognizing the implication of magnetic fields.
The fact that the non-heme iron can be exchanged confirms our prediction about the role played by the iron.
The effect is, however, typically small and decreases proportionally to the inverse square of the wavelength
in an organic material,
although there are isolated reports of large Faraday rotation~\citep{Vandendriessche2012,Vandendriessche2013}.


\item Duplexer (or multiplexer) is the term used in communication engineering, 
whereas physicists use the descriptor time-dependent material and its common name is switch, 
which might function together with other mechanisms, in particular the aforementioned Faraday effect.
Biochemists know that, when a (bacterio)chlorophyll pigment absorbs light, 
it loses an electron to the RC, which already signifies duplexing ~\citep{Berg2015};
when the RC received an electron its $Fe^{3+}$ becomes $Fe^{2+}$ with the magnetism changed.
Although the magnetic field generated by $Fe^{3+}$ is tiny, the subtlety of a switch is that a small change can control an entire device;
in photosynthesis it can lock the incoming electron.
The internal states of the non-heme iron at the RC can serve to control the direction
of the light propagation~\citep{Scheucher2016}.

In terms of mechanics, the photon received might cause the $\alpha$-helical polypeptides/ carotenoid/ bacteriochlorophyll complex
to alter its shape or orientation, 
hence rendering impracticable the flow of the electromagnetic wave in the other direction.
Theorists have previously remarked that the FMO complex might work as a rectifier for unidirectional energy flow without specifying how that action proceeds~\citep{Ishizaki2009}.
Such conformation change is also observed in retinal molecules~\citep{Hayashi2003a}.
With this mechanism the near unity quantum efficiency of conversion is not peculiar because the electron has been locked in~\citep{Romero2014}.

The wave-like energy transfer or the coherence at room temperature observed by previous authors simply signifies a collective motion (mechanical movement) 
instead of quantum coherence~\citep{Engel2007},
as light at this wavelength, relative to the size of the antenna, would be better considered as wave instead of particle;
as the authors also reached similar conclusion latter~\citep{Wilkins2015}.
The functional role played by the persistent spectrometric signal observed simply mean duplexing as discussed here.
The bacteriochlorophylls move in unison with the light wave as seaweed moves with the tides.
A recent finding of photon anti-bunching from LH supports such a scenario,
although no further reason for the physics behind conformational change was provided~\citep{Kruger2011,Wientjes2014}.

Current methods of imaging with a microscope not only are limited by the wavelength used 
but also require harsh conditions or direct mechanical interaction with the samples, 
hence becoming inappropriate for living cells.
Cyro-electron microscopy can certainly be employed~\citep{Frank2018}.
A newly invented imaging method using phonons (acoustic waves) of sub-optical wavelength
or  a photonic crystal-enhanced microscope might enable one to observe the route that a photon takes on entering and leaving LH,
to discover how the LH alters its conformation~\citep{Perez-Cota2016,Zhuo2016}.

\item Thirdly, matamaterial are using sub-wavelength structure, in particular a notched ring~\citep{Enkrich2005}, 
to achieve non-reciprocity, of which the most noticeable future is chirality.
Ullah {\it et al.} use asymmetric particles to achieve chirality of their nanoantenna~\citep{Ullah2018}, while
each module of the bacteria light harvester composed of one carotenoid two short $\alpha$-helical polypeptides, and three bacteriochlorophylls
to achieve chirality. 
If the notched ring is further divided into sub-units the interleaving inductors will guide the flow of displacement current better~\citep{Alu2008a}.

\item The shape of the light harvesters is toroidal as required for anapole radiation.
Toroidal structures can support exotic high-frequency electromagnetic excitations that
are neither electric nor magnetic multipoles, called toroidal moments, which can enhance inter-molecular interaction and energy transfer.
\end{itemize}

The above statements have not excluded a spherical shape, 
which occurs on evoking the Poincar\'e-Brouwer theorem~\citep{Brouwer1911}.
This theorem was originally proven by Poincar\'e and is sometimes called
the hairy-ball theorem; it states that there exists at least one point on the surface of a sphere at which vectors of electric and magnetic fields become equal to zero.
The Poynting vector is also zero at this point. 
This theorem indicates a toroidal shape instead of a spherical circulator, 
which is the second lesson from nature about solar light harvesting mentioned by previous authors~\citep{Scholes2011} and
is consistent with the requirement of anapole radiation,
but is not honoured in many designs of bio-inspired optical antennae~\citep{Hong2015}.
The non-spherical requirement is depicted not only at the scale  for the light-harvesting complex discussed but also
on a larger scale as in chromatophores~\citep{Chandler2009},
although the later authors describe the shape of the chromatophores of {\it Rhodobacter sphaeroides} as spherical in their abstract but as vesicles later in the
content of their paper.

\section{Non-reciprocity of Hqr.}
\label{archaean}

It is natural to ask whether the proposed mechanisms could be equally applicable to archaea?
The answer should be affirmative because anything that mediates between an electromagnetic field and a final sink of the energy can be considered as an antenna;
the only difference is their performance.
There are two other hints indicating that the antenna theory is applicable:
\begin{itemize}
\item
Recently optical antennae configuration
with fourfold ($C_{4v}$) symmetry, a cross-shape antenna consisting of two perpendicular dipole antennae
with a common feed gap, has been considered~\citep{Biagioni2009}. 
\item 
When the cells grow and divide, they do not always separate from each other, 
and form two dimensional arrays of ten or more units like un-separated postage stamps.
\end{itemize}
The cross-shape dark lines on the surface of Hqr. confirm the first point, while
our previously proposed physical mechanism for the bacteria light harvester to coalesce seems to confirm the second point~\citep{Ting2018}.
The cross-shape antenna if made from gold with complex permittivity $\epsilon = -26.64 + 1.66 i $
and illuminated with $800~nm$ circularly polarized light will have best performance at a side length around $170~nm$~\citep{Biagioni2009}.
This is close to our requirement that Hqr. can survive under UV frequency.
The question is how? And, is it a good antenna?
What kind of design does the nature employed to increase the light-harvesting efficiency?
The key to uncover the answer should still in its non-reciprocity.

The light harvesting unit larger than the bacteriorhodopsin is the whole cell for this microbe, 
unlike the bacteria had light harvesters LH1/LH2 inside its cell.
There is no reaction center, and no ion at the reaction center, which appears in bacteria light harvesters and is responsible for
making the light harvesters non-reciprocal.
But the physics still requires a mechanism to make the antenna non-reciprocal.

How could the cell makes the antenna non-reciprocal as previously proposed?
The nonlinearity mechanism proposed seems to be able to serve this purpose:
\begin{itemize}
\item Mahmoud {\it et al.} proposed to manufacture a non-reciprocal artificial metastructure by divide its surface into nano-cells 
with neighbouring areas in disparate chirality~\citep{Mahmoud2015}.
The cross shape appears on the cell wall divide the square into four smaller squares.
\item Some bacteriorhodopsins might form modules to create chirality~\citep{Ullah2018}.
\item Since there have been at least two types of bacteriorhodopsin discovered,
they might also have different chirality and located at different division of the cell-wall of the archaea, i.e.
the trimer might have chirality while the dimer not.
\end{itemize}
These three points indicate that Hqr. is a naturally existing metamaterial~\citep{Zhao2009},
a material that owes its properties to a sub-wavelength structure rather than to its chemical composition~\citep{Pendry2006}.
The bacteriorhodopsin structures analyzed so far are all obtained under non-physiological conditions that have mixed the molecules.

We hence need a study of the structure of Hqr. {\it in situ}, which can be done using a cryo-electron (transmission) microscope,
and which involves a series of experimental steps followed by some software analysis
that does not require the sample to be crystalized and is particular suitable for biological molecules.

Because Hqr. is very thin, it can be accommodated within a slice of the sample prepared, and cool down rapidly by
plunge-freezing to be embedded in amorphous ice~\citep{Tivol2008}.
A bacteriorhodopsin structure, 1BRD, has been solved using cryo-electron microscope~\citep{Henderson1990},
with a resolution of 3.5 \AA~ in a direction parallel to the membrane plane but lower than this in the perpendicular direction.
A single-particle cryo-electron microscope covers from particles of mass 20 kDa up to several megadaltons~\citep{Merk2016}.
The larger is the molecule, the better is the image after averaging.
The images obtained can be analyzed by SPIDER or RELION~\citep{Cheng2015}.
The software will first sort the two-dimensional images into classes according to their orientation
then reconstruct a three-dimensional structure.

\section{Application}
\label{application}

As the geometry is known, we interpret its physical mechanisms that might be useful for the design and production of artificial light-harvesting systems~\citep{Scholes2011,Yu2014,Kruger2016},
optical sensors, wireless power chargers, passive heat radiators, and photocatalytic hydrogen generators:

\begin{itemize}

\item 

An immediate application is to predict what kind of particles should be use to separate sunlight into
a photovoltaic useful range and thermally useful range to enhance the efficiency of solar cells~\citep{Hjerrild2016,Hjerrild2017}.
The nanoparticles that Hjerrild {\it et al.} considered are disk-shaped metallic
particles of diameter roughly $100~ nm$, which exploit plasmon resonances. 
We have shown that the optimal shape of the particles should be
toroidal instead of discotic (or nanorods for the infrared range).  
Hjerrild {\it et al.} attributed the geometry dependence to the role of dipole moments in the plasmon resonance,
while we have shown geometry is itself important.
They wrote that, to avoid scattering, the size of the particles should be limited to $50~nm$, 
whereas the naturally designed light harvester that we considered has a size about $10~nm$.
They further pointed out that silver is the best material to use, but that is quite expensive.  
They considered the material used to be important for the frequencies of absorption, and mentioned that the ratio
of surface to volume of these nanoparticles makes the particles susceptible to damage from high power.  
Such arguments are generally based upon the band-gap theory~\citep{Jain2018}.
But that is for the case of single atom. 
For bulk material time-lag (spatial dispersion) between atoms become important. 
Hence we came to the world of nanoantennae.
Our work shows that a dielectric can be used to avoid such high-power damage, 
which is actually the best material at a scale of $10~nm$,
and that the frequency of absorption can be tailored with the radius of the toroidal particle~\citep{Ting2019}.  
As those authors coated their metallic disk with dielectric, there is no need for a metal at all.
Such nano-particles are also directly used in solar cells for light trapping~\citep{Catchpole2008}.

\item Secondly, we can make optical sensors which are devices that must interact with light strongly.
(Bacterio)rhodopsin and (bacterio)chlorophyll are two molecules in biology that can serve this purpose.
Most animals use rhodopsin, while most plants and bacteria use chlorophyll, except archaea.
As a receiver nanoantennae concentrate light from far-field, enlarging the effective cross section of molecules.
Most bio-mimetic or bio-inspired optical sensors are related to animals~\citep{Martin-Palma2019}.
Plant and virus which surely come to the earth earlier than animals are not discussed yet.

\item Wireless power transfers, presently, works not only under microwave frequencies, if not radio frequencies, 
but also only under designed (non-physiological) condition~\citep{Shinohara2014,Miller2014,Tsai2016}.
They are at most separable transformers, whose coils are not considered as antennae.
However, medical applications require the power chargers to work under flexible physiological condition~\citep{Miller2014}.
The next generation wireless power transfer based upon our design principles will use incoherent light as the nature use, 
which is sustainable, renewable, and working under physiological condition.

\item 

Thermodynamics told us that there are three types of heat transfer, i.e., conduction, convection, and radiation. 
There are situations we have to rely on radiative cooling, such as devices operate in vacuum and
fanless notebook computers.
Radiation is a kind of electromagnetic wave.
If a system emits more electromagnetic waves than it absorbs, it will be cooled down.
Heat radiators normally operate with wavelength between $0.1$ to $100 ~\mu m$~\citep{Modest2013}.

There are two kinds of radiative heat transfer theory discussed in the literature.
The near-field energy transmission, correspond to the F\"orster effect in the photosynthesis community, 
contains photon tunneling other than wave interference when the separation distance is less than the thermal wavelength~\citep{Pendry1999a}.
Such energy transfer is important in photosynthesis because the light received by LH2
needs to find an LH1 to reach the RC and the energy need to migrate from module to module within the antenna to the RC, 
but is of less importance in radiative-cooling because every radiator, which is equivalent to LH2, can function independently. 
On the other hand, the far-field effects are generally considered using Mie theory, 
which describe the particles only as spheres~\citep{Kupiec2010,Johansson2017}.
The radiator, according to the nature design of light harvester, radiate more than Planck's law predicted 
because its characteristic length is smaller than the thermal wavelength~\citep{Fernandez-Hurtado2018}.

\item It can also enhance the efficiency photocatalytic hydrogen generation~\citep{Clarizia2017}.

\item Radiation protective cloaks~\citep{Fleury2015}: 
{\it Deinococcus radiodurans} that is one of the most radiation-resistant organisms known, 
apparently uses two passive mechanisms, other than active DNA repair normally attributed, discussed above~\citep{Levin-Zaidman2003}.
Each cell has two perpendicular furrows that result in a tetrad morphology, which closely resembles Hqr.
Cells of dimer, tetramer and even multimer morphologies can also be obtained.
Such a mechanism has been discussed as forming electromagnetic multipoles to diminish the excess radiation~\citep{Ting2018}.
Its S-layer is a nature made metamaterial.

\end{itemize}

\section{Summary}

In this perspective, we sought physical interpretations/mechanisms 
to consider a light-harvesting antenna as a device to receive electromagnetic waves.
Although theories for light-harvesting in photosynthesis based on classical electrodynamics have been proposed and applied before~\citep{Zimanyi2010,Cleary2013},
antenna theory has never been used to explain light harvesters, even though they have been called antennae for decades~\citep{Parson1974}.
A major theme of the present perspective is that the geometry is more important than the material property, 
which is consistent with what has been found for metamaterials, even though our starting point is antenna theory.
We have shown in this paper that nature uses a metamaterial design in several places.


Our model from classical electrodynamics has the structural information enforced.
We thereby provide explanations for 
\begin{enumerate}
\item the function of the notch,
\item the function of PufX/W,
\item the function of the special pair,
\item the function of the non-heme iron at the LH1-RC complex, 
\item the shape of the LH must not be spherical,
\item the cross section of the light harvester must not be circular,
\item preferring a toroidal shape,
\item an upper-bound for the characteristic length of an efficient light harvester,
\item the usage of dielectric as an antenna,
\item a mechanism to prevent damages from excess sunlight,
\item the dimerization mechanism under intensive light,
\item a mechanism to achieve dual-band radiation spectrum,
\item the wave-like energy transfer observed, 
\item the physics behind photon anti-bunching, 
\item reasons for the modular design of the light harvester, and
\item the reason why each module of the light harvester is composed of one carotenoid two short $\alpha$-helical polypeptides, and three bacteriochlorophylls.
\end{enumerate}
We have discovered also the number of modules involved in each light harvester can be
explained by the language of fractal~\citep{Ting2019c}.
The connectivity matrix thus derived is what we obtained years ago from chemical rate equations~\citep{Ting1999b}.
We show that not only chemistry but also physics illuminate the problem of photosynthesis.

\bibliography{d:/library} 

\end{document}